\documentclass[showpacs,amsmath,amssymb,aps]{revtex4}                                                                                                                                                                                                                                                                                                                                                                                                                                                                                                                                                                                                                                                                                                                                                                                                                                                                                                                                                                                                                                                                                                                                                                                                                                                          
\usepackage{graphicx}
\usepackage{epstopdf}
\usepackage{amsmath}
\usepackage{bm}

\begin{document}

\title{QCD emergent from spontaneous breakdown of relativity}

\author{Kimihide Nishimura}
\email{kmdns@tune.ocn.ne.jp}
\affiliation{Nihon-Uniform, 1-4-1 Juso-Motoimazato, Yodogawa-ku, Osaka 5320028 Japan}

\date{\today}

\begin{abstract}
Spontaneous violation of relativistic invariance of the vacuum  can derive quantum chromodynamics from an U(1) Higgs model including fermions, if the emergent theory is Lorentz invariant. In this model, the vacuum becomes anisotropic, and a fermion created on a triplet of spin-one vacua  acquires degeneracy analogous to the color degrees of freedom. The Nambu-Goldstone bosons originating from the breakdown of rotational symmetry provide the quasi fermions with SU(3)$\times$U(1) effective interactions, which are interpretable as mediated by gluons and photons. 
The confinement of quasi quarks as well as that of Nambu-Goldstone gluons follow from the Lorentz invariance of the emergent theory. 
\end{abstract}

\pacs{12.38.-t, 11.30.Cp, 12.10.-g, 74.20.-z}

\maketitle

1. Recent progress in quantum field theory reveals that spontaneous breakdown of relativity will be responsible for the existence of leptons interacting with weak and electromagnetic forces \cite{KN1,KN2}, as well as for the explanation of the absence of antimatter in the universe. It has been suggested further in \cite{KN1} that quarks and strong interactions may also be derivable in a similar way. This paper proves this conjecture; QCD emerges from spontaneous violation  of rotational symmetry.

The idea of achieving a unified understanding of elementary particles and forces by regarding them as quasi excitations is old, which dates back to the papers of Nambu and Jona-Lasinio \cite{NJ1,NJ2}, appeared four years after the BCS theory of superconductivity \cite{BCS}. However, no realistic models comparable to the standard theory of elementary particles had not been proposed so far. The reason is now clear, since the implementation of the idea requires spontaneous breakdown of Lorentz symmetry.

Although experiments have rigorously confirmed Lorentz invariance of the laws of nature \cite{KR}, those results do not exclude the breakdown of relativistic invariance of the quantum vacuum. 
As presently discussed, spontaneous spacetime symmetry breaking occurs generally when fermions are present in gauge theories in which Higgs fields give mass to the gauge bosons \cite{Higgs1,Higgs2,Higgs3,Englert}. In this case, the vacuum expectation values of fermion currents and gauge potentials break Lorentz symmetry. 
It is worth mentioning, however, that an effective theory emergent from spontaneous breakdown of relativity can still be Lorentz invariant, since the vacuum itself is not directly observable. 
In this view, the actual effect of spontaneous breakdown of Lorentz symmetry in quantum field theory will simply be the rearrangement of field theoretical representation of Lorentz symmetry.

If we  postulate Lorentz invariance of the emergent theory,  the dispersion relation of an emergent fermion should at least satisfy the relation:
\begin{equation}
(p-\delta)^2-m^2=0,
\label{MDR}
\end{equation}
where $\delta^\mu$ is a constant 4-vector. 
This type of dispersion relation will not be conflict with relativity, since CPT-violating term $\delta^\mu$ is absorbable by a suitable phase transformation applied to the effective spinor field \cite{CK}. 
Nevertheless, the existence of non-zero $\delta^\mu$ does affect the energy density of the system. 
When $\delta^\mu$ is spacelike, the spacial vector $\bm{\delta}$ can not vanish in any reference frame, and the vacuum breaks rotational symmetry in any Lorentz frame. In this case, we can expand the vacuum into a series of spherical harmonics, and correspondingly, a state of a fermion on the broken vacuum can be expressible as a superposition of the fermion states on the component vacua. 

Though, at first sight, these states seem to represent fermions with higher spins, this is not the unique interpretation, or rather  we should adopt a different interpretation in view of Lorentz invariance of the emergent theory. For example, if we consider the spin-one vacua, degenerating triply, the general basis for the triplet has maximally U(3)$=$SU(3)$\times$U(1) symmetries, and the corresponding fermion states on them will also participate in the same symmetry transformations. Consequently, a fermion acquires three degrees of degeneracy corresponding to the color charges in QCD \cite{FGL}, as well as  an U(1) charge analogous to the electricity in QED.  
We remark here that the emergent symmetries become larger than the broken ones, since mathematically each multiplet of vacua spans a complex vector space.

The bases of vacuum multiplets can differ at different spacetime points. Then the global symmetries will be extended to the  local ones.
As the nature of spontaneous symmetry breaking, the effective  interactions corresponding to these emergent local symmetries are expected to be mediated by the Nambu-Goldstone vector mesons \cite{Goldstone,GSW}. We will present also a formal method to derive the Lorentz invariant effective Lagrangian for these Nambu-Goldstone vector mesons, consisting of their kinetic terms and the interaction terms with quasi quarks.

2. We first exemplify by using an U(1) Higgs model with a massless Dirac fermion that the occurrence of spontaneous  Lorentz violation can further stabilize the vacuum.  
The Higgs coupling to the Dirac fermion is not considered, since the fermion mass is inessential for the following argument. 
The Lagrangian is given by
\begin{equation}
\begin{array}{lcl}
{\cal L}&=&
D^\mu\phi^\dagger D_\mu\phi
-\frac{1}{4}Y^{\mu\nu}Y_{\mu\nu}
+\bar{\psi}i\gamma^\mu D_\mu\psi={\cal T}-{\cal V},\\
{\cal T}&=&\bar{\psi}i\gamma^\mu\partial_\mu\psi
-\frac{1}{4}Y^{\mu\nu}Y_{\mu\nu}
+\partial^\mu\phi^\dagger \partial_\mu\phi,\\
{\cal V}&=&j^\mu Y_\mu
-g^2Y^\mu Y_\mu\phi^\dagger\phi
+\frac{\lambda}{4}(\phi^\dagger\phi-\eta^2)^2,\\
j^\mu&=&j^\mu_\psi+j^\mu_\phi,\\
j^\mu_\psi&=&g\bar{\psi}\gamma^\mu\psi,\\
j^\mu_\phi&=&ig(\phi^\dagger \partial^\mu\phi-\partial^\mu\phi^\dagger\phi),\\
\label{LagU(1)}
\end{array}
\end{equation}
where $\psi$ is a Dirac spinor, $Y^{\mu\nu}$ is the field strength of an U(1) gauge field $Y^\mu$, and $\phi$ is a complex scalar field with the Higgs potential. We have taken for simplicity the same coupling constant $g$ in the covariant derivative: $D_\mu=\partial_\mu+igY_\mu$ either for the Dirac  field and for the Higgs field.
Assuming the total gauge current: $j^\mu=j^\mu_\psi+j^\mu_\phi$ has the vacuum expectation value $\langle j^\mu\rangle$, we minimize the potential density ${\cal V}$ with respect to $\phi$ and $Y^\mu$ to find the extremal conditions:
\begin{equation}
\begin{array}{cc}
\langle j^\mu\rangle=2g\phi_0^2\delta^\mu,&\phi_0^2=\eta^2+2\delta\cdot\delta/\lambda, \\
\label{SCC}
\end{array}
\end{equation}
where $\phi_0=\sqrt{\langle\phi^\dagger\phi\rangle}$ and 
$\delta^\mu=g\langle Y^\mu\rangle$.
We have excluded solution $\phi_0=0$ from (\ref{SCC}), which  corresponds to the maximum of the Higgs potential. 
Rewriting $Y^\mu$ in ${\cal L}$ by 
$gY^\mu=\delta^\mu+gZ^\mu$, while $\phi$ by
$\phi=e^{i\theta}(\phi_0+\sigma/\sqrt{2})$ with real field $\sigma$, 
we obtain, after eliminating the phase factor $e^{i\theta}$ by a suitable gauge transformation, for the mass of the neutral weak boson $m_Z$ and for the mass of the Higgs boson $m_H$
\begin{equation}
\begin{array}{cc}
m_Z^2=2g^2\eta^2+4g^2\delta\cdot\delta/\lambda,&m_H^2=\lambda\eta^2+2\delta\cdot\delta,
\end{array}
\label{M_ZM_H}
\end{equation}
where the $\phi_0$-dependence has been eliminated from (\ref{M_ZM_H}) by using (\ref{SCC}).
The effective Lagrangian for the free Dirac fermion then becomes\begin{equation}
{\cal L}_\psi=\bar{\psi}\gamma^\mu( i\partial_\mu-\delta_\mu)\psi,
\label{EDL}
\end{equation}
and the extremum of ${\cal V}$ is given by
\begin{equation}
{\cal V}_{\rm ext}=\eta^2\delta\cdot\delta\left(1+3\frac{\delta\cdot\delta}{\lambda\eta^2}\right).
\label{V_ext}
\end{equation}
As mentioned earlier, we see from (\ref{V_ext}) that the constant 4-vector $\delta^\mu$ does affect the potential energy density of the system, though it could have been  absorbed by some gauge transformation. In the case of ordinary Higgs mechanism, we have $\delta^\mu=0$, and thereby ${\cal V}_{\rm ext}=0$.
However, ${\cal V}$ can become even negative when $\delta^\mu$ is a spacelike vector and its magnitude lies in the range: $0<-\delta\cdot\delta<\lambda\eta^2/3$.  Accordingly, if fermions are present, the breakdown of symmetries caused by the Higgs field can not stop within gauge symmetries, but proceeds even to spacetime symmetries. 

We further find that $\langle j^\mu_\phi\rangle=0$, while $\langle j_\psi^\mu\rangle$ is expressible as $g\Gamma\delta^\mu$. The Lorentz covariant estimation of $\Gamma$ by using the propagator for the quasi fermion  described by (\ref{EDL}) gives
\begin{equation}
\begin{array}{cc}
\langle j^\mu\rangle=-{\rm Tr}\gamma^\mu S_F(0)=-4gI_1^\mu,&
I_1^\mu=\displaystyle\int\frac{d^4p}{(2\pi)^4}i\frac{(p-\delta)^\mu}{(p-\delta)^2+i\epsilon}.
\end{array}
\label{S_F(0)&I_1}
\end{equation}  
By expanding $I_1^\mu$ with respect to $\delta^\mu$, we find
\begin{equation}
\begin{array}{cc}
I_1^\mu=-\displaystyle\frac{\delta^\mu}{2}\left(k_1+\frac{\delta\cdot\delta}{48\pi^2}\right),
k_n=\displaystyle\int\frac{d^4p}{(2\pi)^4}\frac{i}{[p^2+i\epsilon]^n},
\end{array}
\label{IK}
\end{equation}  
from which we obtain 
\begin{equation}
\Gamma=2k_1+\displaystyle\frac{\delta\cdot\delta}{24\pi^2},
\label{Gammas}
\end{equation}  
where the Lorentz-invariant constant $k_1$ with mass dimension 2 is quadratically divergent.
We find from (\ref{SCC}) and (\ref{Gammas}) that
\begin{equation}
\delta\cdot\delta=\displaystyle\frac{\lambda}{2}
\frac{k_1-\eta^2}
{1-\displaystyle\frac{\lambda}{96\pi^2}}.
\label{delta^2}
\end{equation}  
When $\delta\cdot\delta<0$, the following inequality holds:
\begin{equation}
\displaystyle\frac{1}{3}+\frac{\lambda}{144\pi^2}<\frac{k_1}{\eta^2}<1.
\label{Inequality}
\end{equation}  
If spontaneous spacetime symmetry breaking does not occur, we have $\phi_0=\eta$, and $k_1=\eta^2$. Even otherwise, the inequality (\ref{Inequality}) still requires that the quadratic divergence $k_1$ and the parameter $\eta^2$ in the Higgs potential should have the same order of magnitude.
This observation contrasts to the result from the dimensional regularization \cite{THV}, in which $k_1$ is put equal to zero. We will come back to this  point later.

3. We have seen that when $\delta^\mu$ is spacelike, the relativistic invariance of the vacuum can be broken spontaneously. 
In order to show that the quasi fermion emergent in this case can acquire triple degeneracy, which is the  analogue of the color charges of quarks, we henceforth discuss in a Lorentz frame in which $\delta^\mu=(0,\bm{\delta})$, and confirm first the anisotropy of the vacuum by observing the relation:
\begin{equation}  
\begin{array}{cc}   
\langle\Omega\vert i[L_\theta, \bm{j}]\vert\Omega\rangle=\bm{\theta}\times\langle\bm{j}\rangle\neq\bm{0},&L_\theta=\bm{\theta}\cdot\bm{L},
\end{array}
\label{VacAns}
\end{equation}
for $\bm{\delta}\neq\bm{0}$, where $\bm{L}$ is the generator of rotations, from which follows $L_\theta\vert\Omega\rangle\neq0$.
The anisotropic vacuum is expressible as a superposition of various vacua with different spins, which is obtained by expanding $|\Omega\rangle$ in a series of spherical harmonics 
$|\Omega_{lm}\rangle$ with respect to $\bm{L}$: 
\begin{eqnarray}
\vert\Omega\rangle=\sum_{lm}c_{lm}|\Omega_{lm}\rangle,&
\bm{L}^2|\Omega_{lm}\rangle=l(l+1)|\Omega_{lm}\rangle,&
L_3|\Omega_{lm}\rangle=m|\Omega_{lm}\rangle.
\label{AVExp}
\end{eqnarray}

On the other hand, the quasi fermion described by Lagrangian (\ref{EDL}) has 4-momentum $p^\mu=(\pm\vert\bm{p}-\bm{\delta}\vert,\bm{p})$. 
We here adopt the ``hole" interpretation to the negative energy states; the absence of a quasi fermion with  4-momentum $p^\mu$ is regarded as the presence of a quasi antifermion with 4-momentum $-p^\mu$. Then the quasi antifermion has four-momentum $p^\mu=(\vert\bm{p}+\bm{\delta}\vert,\bm{p})$. We next introduce annihilation operators $q_{\bm{p}s}$ and $\bar{q}_{\bm{p}s}$ for a quasi fermion and a quasi antifermion with 3-momentum $\bm{p}$ and spin $s$, respectively, which satisfy anti-commutation relations: $\{q_{\bm{p}s},q_{\bm{p}'s'}^\dagger\}=\{ \bar{q}_{\bm{p}s}, \bar{q}_{\bm{p}'s'}^\dagger\}=\delta_{ss'}\delta_{\bm{p}\bm{p}'}$, while those for the other combinations vanish.
As the result, the quantized Dirac spinor $\psi$ can be expanded as
\begin{equation}
\psi(x)=
\displaystyle\sum_{\bm{p}s}
q_{\bm{p}s}u_{\bm{p}s}(x)+\bar{q}^\dagger_{\bm{p}s}v_{\bm{p}s}(x),
\label{QQPsi}
\end{equation}
where $u_{\bm{p}s}(x)$ and $v_{\bm{p}s}(x)$ are appropriately normalized eigenfunctions corresponding to positive and negative energies, respectively.  
In these notations, the vacuum state $|\Omega\rangle$ can be characterized by the conditions: $q_{\bm{p}s}|\Omega\rangle=\bar{q}_{\bm{p}s}|\Omega\rangle=0$ for arbitrary $\bm{p}$ and $s$.

We now examine the rotational invariance of the emergent theory by considering a one-particle state of anisotropic quasi fermion: $q_{\bm{p}s}^\dagger\vert\Omega\rangle$.  
As already mentioned, this fermion would not behave as a particle with spin $1/2$, even if an appropriate phase transformation can remove the constant spacial vector $\bm{\delta}$ from the effective equation of motion, since the vacuum $\vert\Omega\rangle$ is not rotationally invariant. 

If one-particle state $q_{\bm{p}s}^\dagger\vert\Omega\rangle$ is expressed as 
\begin{eqnarray}
q_{\bm{p}s}^\dagger\vert\Omega\rangle=\sum_{lm}c_{lm}q_{\bm{p}s}^\dagger\vert\Omega_{lm}\rangle, 
\label{1PSE}
\end{eqnarray}
one of the component states $q_{\bm{p}s}^\dagger\vert\Omega_{lm}\rangle$ can further be rewritten as a superposition of fermion states, each of which has spin $s$ in the range $\vert l-1/2\vert\leq s\leq l+1/2$, and therefore fermions with spin higher than $1/2$ seem to emerge when Lorentz symmetry is spontaneously broken.
However, this interpretation contradicts the hypothesis that the emergent theory should be Lorentz invariant, since the detection of a fermion with spin higher than $1/2$ implies an indirect evidence of anisotropy of the vacuum, or the breakdown of rotational symmetry. 
Accordingly, the detection of an anisotropic quasi fermion (\ref{1PSE}) is considered as forbidden by our hypothesis.  

Nevertheless, this restriction does not exclude the appearance of a composite particle, which is made of anisotropic quasi fermions, 
if they form such a state in which the contributions from  component vacua to its rotational property cancel in total.
This possibility concerns the emergence of symmetries larger than the broken ones for composite particles. 

We concentrate on the vacua with spin one:  
$|\Omega_{1m}\rangle$ in (\ref{AVExp}), and define a new  triplet $\vert\Omega_i\rangle$ satisfying  
\begin{equation}
\begin{array}{cc}
\langle\Omega_i\vert\Omega_j\rangle=\delta_{ij},&
L_i\vert\Omega_j\rangle=i\epsilon_{ijk}\vert\Omega_k\rangle,
\end{array}
\end{equation}
by the relations:
\begin{equation}
\begin{array}{ccc}
\vert\Omega_1\rangle=\displaystyle\frac{1}{\sqrt{2}}
\left(\vert\Omega_{11}\rangle-\vert\Omega_{1-1}\rangle\right),&
\vert\Omega_2\rangle=\displaystyle\frac{1}{\sqrt{2}i}
\left(\vert\Omega_{11}\rangle+\vert\Omega_{1-1}\rangle\right),  &   
\vert\Omega_3\rangle=\vert\Omega_{10}\rangle.  
\end{array}
\end{equation}
The new triplet $\vert\Omega_i\rangle$ transforms the same as a real spacial vector under rotation. 
However, since $\vert\Omega_i\rangle$ are quantum states, we  may redefine the basis for spin-one vacua by using their  superposition with complex coefficients: 
\begin{eqnarray}
\vert\Omega_\alpha\rangle=\sum_ic_{\alpha i}^*\vert\Omega_i\rangle&{\rm or}&
\vert\bar{\Omega}_\alpha\rangle=\sum_ic_{\alpha i}\vert\Omega_i\rangle,
\end{eqnarray}
where $c_{\alpha i}$ satisfies $\sum_i c_{\alpha i}c^*_{\beta i}=\delta_{\alpha\beta}$ and $\sum_\alpha c_{\alpha i}c^*_{\alpha j}=\delta_{ij}$. 
Then an arbitrary U(3) transformation: $\vert\Omega'_\alpha\rangle=\sum_\beta u^*_{\alpha\beta}\vert\Omega_\beta\rangle$, or $\vert\bar{\Omega}'_\alpha\rangle=\sum_\beta u_{\alpha\beta}\vert\bar{\Omega}_\beta\rangle$ preserves the orthonormal conditions: 
\begin{equation}
\begin{array}{cc}
  \langle\Omega'_\alpha\vert\Omega'_\beta\rangle=\delta_{\alpha\beta},&
\langle\bar{\Omega}'_\alpha\vert\bar{\Omega}'_\beta\rangle=\delta_{\alpha\beta}.
\end{array}
\end{equation}
We then reconsider one-particle states $\vert q_{\alpha n}\rangle=q_n^\dagger\vert\Omega_\alpha\rangle$ and 
$\vert \bar{q}_{\alpha n}\rangle=\bar{q}_n^\dagger\vert\bar{\Omega}_\alpha\rangle$,  
where momentum and spin indices $(\bm{p},s)$ have been  abbreviated to a single index $n$, for simplicity.
The antifermion should be created on $\vert\bar{\Omega}_\alpha\rangle$ due to the consistency under U(3) transformations. We further change notations by transferring the index $\alpha$ of spin-one vacua to the $q$-operator according to the following rules:
\begin{equation}
\begin{array}{cc}
q_n^\dagger\vert\Omega_\alpha\rangle=q_{\alpha n}^\dagger\vert0\rangle,
 &
\bar{q}_n^\dagger\vert\Omega_\alpha\rangle=\bar{q}_{\alpha n}^\dagger\vert0\rangle,
\end{array}
\label{q_alpha}
\end{equation}
where $\vert0\rangle$ is the symbolic state introduced as the Lorentz invariant effective vacuum. 
Similarly, a many-body state expressed as the direct product of one-particle states is transfered as
\begin{equation}
\vert q_{\alpha n}\rangle\cdots\vert q_{\gamma n'}\rangle
\vert \bar{q}_{\delta m}\rangle\cdots\vert \bar{q}_{\lambda m'}\rangle
=\left(q_{\alpha n}^\dagger \cdots q_{\gamma n'}^\dagger \bar{q}_{\delta m}^\dagger\cdots \bar{q}_{\lambda m'}^\dagger\right)\vert0\rangle,
\end{equation}
where two one-particle states anti-commute with each other:
$\vert q_{\alpha n}\rangle\vert q_{\beta m}\rangle=-\vert q_{\beta m}\rangle\vert q_{\alpha n}\rangle$. 
 Under these premises an U(3) transformation for quasi particle states is also transferrable to the effective operators $q_{\alpha n}$, $\bar{q}_{\alpha n}$, and to the effective vacuum $\vert0\rangle$ as 
\begin{eqnarray}
q'_{\alpha n}=Uq_{\alpha n}U^{-1}=\displaystyle\sum_\beta u_{\alpha\beta}q_{\beta n},&\bar{q}'_{\alpha n}=U\bar{q}_{\alpha n}U^{-1}=\sum_\beta u^*_{\alpha\beta}\bar{q}_{\beta n},& U\vert0\rangle=\vert0\rangle.
\end{eqnarray}
It will be clear in these notations that a totally antisymmetric state of three quasi fermions on different spin-one vacua: 
\begin{equation}
\vert q_{n_1}q_{n_2}q_{n_3}\rangle
=\frac{1}{\sqrt{6}}\sum_{\alpha\beta\gamma}\epsilon_{\alpha\beta\gamma}
\vert q_{\alpha{n_1}}\rangle\vert q_{\beta {n_2}}\rangle\vert  q_{\gamma {n_3}}\rangle
=\frac{1}{\sqrt{6}}\sum_{\alpha\beta\gamma}\epsilon_{\alpha\beta\gamma}
q_{\alpha{n_1}}^\dagger q_{\beta{n_2}}^\dagger q_{\gamma {n_3}}^\dagger\vert0\rangle,
\label{BS}
\end{equation}
is invariant under SU(3) transformations, and therefore a three-body state of this type is devoid of extra spins coming from anisotropy of the vacuum. Accordingly, each constituent quasi fermion may  be regarded as an ordinary spin $1/2$ relativistic particle.  
In addition, a two-body state:  
\begin{equation}
\vert q_{n_1}\bar{q}_{n_2}\rangle
=\frac{1}{\sqrt{3}}\sum_\alpha
\vert q_{\alpha{n_1}}\rangle\vert\bar{q}_{\alpha{n_2}}\rangle=\frac{1}{\sqrt{3}}\sum_\alpha
q_{\alpha{n_1}}^\dagger \bar{q}_{\alpha{n_2}}^\dagger\vert0\rangle,
\label{MS}
\end{equation}
is similarly invariant under SU(3) transformations and devoid of extra spins from the vacua. In this case (\ref{MS}) is also invariant under U(1) transformations. 

The above arguments show that the anisotropy of the vacuum cancels in three body states (\ref{BS}) or in two-body states (\ref{MS}), which proves that the appearance of these composite particles does not contradict Lorentz invariance of the emergent  theory. 
If we regard (\ref{BS}) as a baryon, while (\ref{MS}) as a meson, then $\alpha$ is identified with the color index for quarks.
This identification implies that, in our quasi fermion picture of quarks, the color degrees of freedom is not inherent in quasi excitations, but originates intrinsically from the vacua on which quasi fermions are excited.

4. The effective spinor operator for quasi quarks in the interaction representation $\psi_\alpha(x)$ is obtainable by replacing $q_{\bm{p}s}$ and $\bar{q}^\dagger_{\bm{p}s}$ in (\ref{QQPsi}) with $q_{\alpha\bm{p}s}$ and $\bar{q}^\dagger_{\alpha\bm{p}s}$, respectively, which satisfies the equation of motion: $\gamma^\mu( i\partial_\mu-\delta_\mu)\psi_\alpha=0$. 
A global U(3) transformation of $\psi_\alpha(x)$, induced by a redefinition of the basis for spin-one vacua, is then expressed as
\begin{equation}
\psi'_\alpha(x)=U\psi_\alpha(x)U^{-1}=\sum_\beta u_{\alpha\beta}\psi_\beta(x).
\end{equation}
This global U(3) transformation will be generalized to a local one, if we take different orthonormal bases $\vert\Omega_\alpha\rangle$ for different spacetime points. Then U(3)=SU(3)$\times$U(1) gauge interactions will be introduced into quasi quarks. 
In the context of spontaneous breakdown of rotational symmetry, corresponding gauge bosons are viewed as the Nambu-Goldstone bosons.  
We next propose a formalism which derives an effective Lagrangian for ``quasi gluons" and ``quasi photons", which have their own dynamics and interact with quasi quarks. If the triplet of spacetime-dependent spin-one vacua is denoted by $\vert\Omega_\alpha(x)\rangle$, two triplets $\vert\Omega_\alpha(x)\rangle$ and $\vert\Omega_\alpha(x')\rangle$ at different spacetime points are transformable by some U(3) transformation matrix $u_{\alpha\beta}(x',x)$ as
\begin{equation}
\vert\Omega_\alpha(x')\rangle=\sum_\beta u_{\alpha\beta}(x',x)\vert\Omega_\alpha(x)\rangle.
\end{equation}
In particular, for two points an infinitesimal distance apart, we have 
\begin{equation}
\vert\Omega_\alpha(x^\mu+dx^\mu)\rangle-\vert\Omega_\alpha(x)\rangle=-idx^\mu\omega_{\mu\alpha\beta}(x)\vert\Omega_\beta(x)\rangle,
\label{VC}
\end{equation}
where connection one-form $dx^\mu\omega_{\mu\alpha\beta}(x)$ is hermitian due to the normalization condition: $\langle\Omega_\alpha(x)\vert\Omega_\beta(x)\rangle=\delta_{\alpha\beta}$. 
If we define the covariant derivative $D_\mu$ on $\vert\Omega_\alpha(x)\rangle$ by 
$(D_\mu)_{\alpha\beta}=\delta_{\alpha\beta}\partial_\mu+i\omega_{\mu\alpha\beta}$, then (\ref{VC}) reads
\begin{equation}
D_\mu\vert\Omega_\alpha(x)\rangle=0,
\label{VCF}
\end{equation}
which implies that the local vacua $\vert\Omega_\alpha(x)\rangle$ are invariant under this covariant derivative.
The triplet of effective Dirac spinors $\psi_\alpha(x)$ in this case is generalized by the relations:
\begin{equation}
\begin{array}{cc}
\psi_\alpha(x)\vert0\rangle=\psi(x)\vert\Omega_\alpha(x)\rangle,
&\psi^\dagger_\alpha(x)\vert0\rangle=\psi^\dagger(x)\vert\Omega_\alpha(x)\rangle.
\end{array}
\label{DTT}
\end{equation}
Since $D_\mu\psi(x)=\partial_\mu\psi(x)$ and $D_\mu\vert0\rangle=0$, 
we obtain from the covariant derivative of (\ref{DTT})  
\begin{equation}
\left[D_\mu\psi_\alpha(x)\right]\vert0\rangle=
\left[\partial_\mu\psi(x)\right]\vert\Omega_\alpha(x)\rangle.
\end{equation}
We may apply this rule to the Lagrangian (\ref{EDL}) to obtain
\begin{equation}
\sum_\alpha\langle\Omega_\alpha(x)\vert{\cal L}_\psi\vert\Omega_\alpha(x)\rangle=\sum_{\alpha\beta}\langle0\vert\bar{\psi}_\alpha i\gamma^\mu [(D_\mu)_{\alpha\beta}-\delta_\mu\delta_{\alpha\beta}]\psi_\beta\vert0\rangle.
\end{equation}
As the result, we have the Lorentz invariant effective Lagrangian for the quasi quarks: 
\begin{equation}
{\cal L}_q=\sum_{\alpha\beta}\bar{\psi}_\alpha i\gamma^\mu  [(D_\mu)_{\alpha\beta}-\delta_\mu\delta_{\alpha\beta}]\psi_\beta,
\label{Lq}
\end{equation} 
with respect to the Lorentz-invariant vacuum $\vert0\rangle$. 
If gauge connection $\omega_{\mu\alpha\beta}$ in (\ref{VC})  is expressed as 
\begin{equation}
\omega_{\mu\alpha\beta}=g'A^a_\mu(t_a)_{\alpha\beta}+g''B_\mu\delta_{\alpha\beta},
\end{equation}
where $t_a$ $(a=1,\cdots,8)$ are  SU(3) generators in the fundamental representation, then $A^a_\mu$ are identified with  SU(3) gauge fields with coupling constant $g'$, while $B_\mu$ with an U(1) gauge field with coupling constant $g''$.

Though, at this stage, $A^a_\mu$ and $B_\mu$ are auxiliary fields which make the effective Lagrangian (\ref{Lq}) to be invariant under SU(3)$\times$U(1) gauge transformations, it has been known that they acquire dynamics by renormalization  \cite{Bjorken, Guralnik,Eguchi}. In fact, the path integral for Lagrangian (\ref{Lq}) gives the effective Lagrangian $\Delta{\cal L}_\omega$ for $A^a_\mu$ and $B_\mu$:
\begin{equation}
\int d^4x\Delta{\cal L}_\omega
=-i{\rm Tr}\ln[1-\gamma\cdot\omega(i\gamma\cdot\partial)^{-1}]
=\int d^4x\left[\ -k_1{\rm tr}\omega^\mu \omega_\mu+\frac{k_2}{3}{\rm tr}\omega^{\mu\nu}\omega_{\mu\nu}
+{\rm finite\ terms}\ \right],
\label{DLF}
\end{equation}
where $\omega_{\mu\nu}=\partial_\mu\omega_\nu-\partial_\nu\omega_\mu+i[\omega_\mu, \omega_\nu]$. The logarithmically divergent constant $k_2<0$ is defined in (\ref{IK}), and ``Tr" is taken with respect to both the spacetime coordinates and the group indices, while ``tr" is taken with respect to the group indices only. 
The ``finite terms" would diverge if evaluated without an infrared cutoff. 
The second term in the last parentheses in (\ref{DLF}) is interpretable as the kinetic terms for $A^a_\mu$ and $B_\mu$ with proper sign,  while the first represents the negative mass term for them.
Since $\omega_\mu$ represents the massless Nambu-Goldstone vector mesons, the negative mass term in $\Delta{\cal L}_\omega$ should be cancelled by some counter term, or $k_1$ should be treated as zero.
Though we do not propose here some mathematical device to provide the mass counter term, it should be remarked that treating $k_1$ as zero, in conformity with the argument of dimensional regularization, does not contradict the statement: $k_1\sim\eta^2$ mentioned earlier, since $\omega_\mu$ is not a fundamental gauge field, but an effective representation of Nambu-Goldstone mesons emergent from the anisotropy of the vacuum, where the contribution from the quadratic divergence $k_1$ has been already taken into account by (\ref{Gammas}). 
Therefore, the appearance of negative mass term in (\ref{DLF}) will be due to the double counting of quadratic divergence.  
From the above considerations, we obtain as the kinetic terms for the effective SU(3)$\times$U(1) gauge fields:
\begin{equation}
{\cal L}_G=-\frac{1}{4}\bm{A}^{\mu\nu}\cdot\bm{A}_{\mu\nu}-\frac{1}{4}B^{\mu\nu}B_{\mu\nu},
\end{equation}
where we have taken ${g'}^2=-3/(2k_2)$ and ${g''}^2=-1/(4k_2)$.
At this stage, constant 4-vector $\delta^\mu$ will be  completely removable from the emergent theory by a suitable U(1) gauge transformation for $B_\mu$.

We recognize here the emergent symmetries higher than the broken one, as well as the Nambu-Goldstone vector bosons more than expected from the Goldstone theorem. 
Concerning these results, which seems to be contrary to conventional expectations, it may be worth mentioning that the emergent SU(3) symmetry is realized only among quasi quarks, which are not directly observable, while the observation of  corresponding Nambu-Goldstone gluons is also forbidden due to  Lorentz invariance of the effective theory.

5. As we have clarified, spontaneous violation of relativity is not a rare phenomenon in quantum field theory. It will be rather  inevitable when the Higgs mechanism operates in a system including fermions.
Nevertheless, experiments have observed no phenomenon  indicating a clear evidence for the violation of Lorentz symmetry. 
These observations lead us to the hypothesis that the effective theory emergent from spontaneous Lorentz violation is still Lorentz invariant. 
We showed as the consequence of this hypothesis that QCD can emerge owing to the anisotropy of the vacuum. The color charge of quarks originates from the triple degeneracy of spin-one vacua. 

We have also seen that the quadratic divergence $k_1$ becomes nearly equal to the parameter $\eta^2$ included in the Higgs potential, while logarithmic divergence $k_2$ relates to the fine structure constant. 
These observations suggest that the concept of spontaneous  breakdown of relativity will give important clues to understand the structure of the standard theory of elementary particles, as well as to solve the divergence problems in quantum field theories.

\end{document}